\documentclass[final,3p,times]{elsarticle}

\usepackage{subfig}
\usepackage{lineno,hyperref}
\usepackage{textcomp}
\usepackage{xcolor}
\usepackage{multirow}
\modulolinenumbers[5]

\journal{Physica Medica}

\begin{document}

\begin{frontmatter}

\title{A Geant4 Based Simulation Platform of the HollandPTC R\&D Proton Beamline for Radiobiological Studies}

\author[myaff1]{C. F. Groenendijk}
\address[myaff1]{Radiation Science \& Technology, Delft University of Technology, Delft, The Netherlands}

\author[myaff2]{M. Rovituso}
\address[myaff2]{Research and Development, Holland Proton Therapy Centre, Delft, The Netherlands}

\author[myaff1]{D. Lathouwers}

\author[myaff3,myaff1]{J. M. C. Brown}
\cortext[mycorrespondingauthor]{Corresponding author: jmbrown@swin.edu.au}
\address[myaff3]{Optical Sciences Centre, Department of Physics and Astronomy, School of Science, Swinburne University of Technology, Melbourne, Australia}

\begin{abstract} 

A Geant4 based simulation platform of the Holland Proton Therapy Centre (HollandPTC, Netherlands) R\&D beamline (G4HPTC-R\&D) was developed to enable the planning, optimisation and advanced dosimetry for radiobiological studies. It implemented a six parameter non-symmetrical Gaussian pencil beam surrogate model to simulate the R\&D beamline in both a pencil beam and passively scattered field configuration. Three different experimental proton datasets (70 MeV, 150 MeV, and 240 MeV) of the pencil beam envelope evolution in free air and depth-dose profiles in water were used to develop a set of individual parameter surrogate functions to enable the modelling of the non-symmetrical Gaussian pencil beam properties with only the ProBeam isochronous cyclotron mean extraction proton energy as input. This refined beam model was then benchmarked with respect to three independent experimental datasets of the R\&D beamline operating in both a pencil beam configuration at 120 and 200 MeV, and passively scattered field configuration at 150 MeV. It was shown that the G4HPTC-R\&D simulation platform can reproduce the pencil beam envelope evolution in free air and depth-dose profiles to within an accuracy on the order of $\pm$5\% for all tested energies, and that it was able to reproduce the 150 MeV passively scattered field to the specifications need for clinical and radiobiological applications.

\end{abstract}


\end{frontmatter}



\section{Introduction}

Proton radiotherapy is a cancer treatment modality that has seen increased use over the last two decades due to its ability to improve local conformity of radiation dose to target tumours whilst sparing the surrounding healthy tissue \cite{Cozzi2001,Palm2007,McGowan2013,Widder2016,Paganetti2018}. The Holland Proton Therapy Centre (HollandPTC) is one of three proton radiotherapy centres in the Netherlands and has treated cancer patients that qualify for proton radiotherapy since 2018 \cite{HollandPTC2018}. HollandPTC is a ProBeam (Varian, a Siemens Healthineers Company) isochronous cyclotron-based facility that features pencil beam scanning into two gantry rooms, one fixed horizontal eye treatment beamline, and one fixed horizontal Research \& Development (R\&D) beamline capable of producing proton energies ranging from 70 MeV up to 250 MeV \cite{Rovituso2023}. HollandPTC is the only proton radiotherapy facility in the Netherlands that possesses a dedicated clinical R\&D beamline, and therefore maximising the scientific output from each beamtime session is of great importance. 

The HollandPTC R\&D beamline, designed and developed specifically for HollandPTC, consists of equipment to build a passive scattering system in order to produce large fields of varying sizes to facilitate radiobiological studies with clinically relevant proton energies \cite{Rovituso2023}. To ensure that every allocated experimental beamtime is used efficiently significant planning, workflow optimsation, and pre-irradiation preparation is required. One useful way to ensure that each experimental configuration will yield the desired proton field shape, intensity, and incident energy spectra is to undertake \textit{in silico} trials through the use of Monte Carlo radiation transport modelling toolkits such as Geant4 \cite{G42003, G42006, G42016}, FLUKA \cite{FLUKA2005,FLUKA2014} and MCNP \cite{MCNP2007,MCNP2014}. This approach is standard for lower energy proton beamlines \cite{Scampoli2001,Constanzo2014,Dahle2017,BR2020,Straticiuc2022}, and has been shown to be crucial for clinical energy proton passive scattering beamlines to enable accurate determination of delivered dose/LET in radiobiological studies \cite{Tommasino2019,Nomura2021}. 

This work presents the development of a Geant4 based simulation platform of the HollandPTC R\&D beamline (G4HPTC-R\&D) to enable the planning, optimisation, and advanced dosimetry for radiobiological studies in both a pencil beam and passively scattered field configuration. In contrast to past studies, this work implements a six parameter non-symmetrical Gaussian pencil beam surrogate model to simulate the pencil beam properties with only the ProBeam isochronous cyclotron mean extraction proton energy as input. Section \ref{sec:exp-methods} describes the development of the G4HPTC-R\&D simulation platform and its individual parameter surrogate functions for non-symmetrical Gaussian pencil beam model through optimisation with respect to three proton energy experimental datasets, and G4HPTC-R\&D's benchmarking with respect to an additional three independent experimental datasets. The results of this process and an accompanying discussion can be found in Section \ref{sec:exp-results}, with an overall conclusion following in Section \ref{sec:diss-con}. 

\section{Methods \label{sec:exp-methods}}

\subsection{G4HPTC-R\&D Simulation Platform Development and Proton Pencil Beam Model Optimisation}

\begin{figure}
    \centering
    \includegraphics[width=0.70\textwidth]{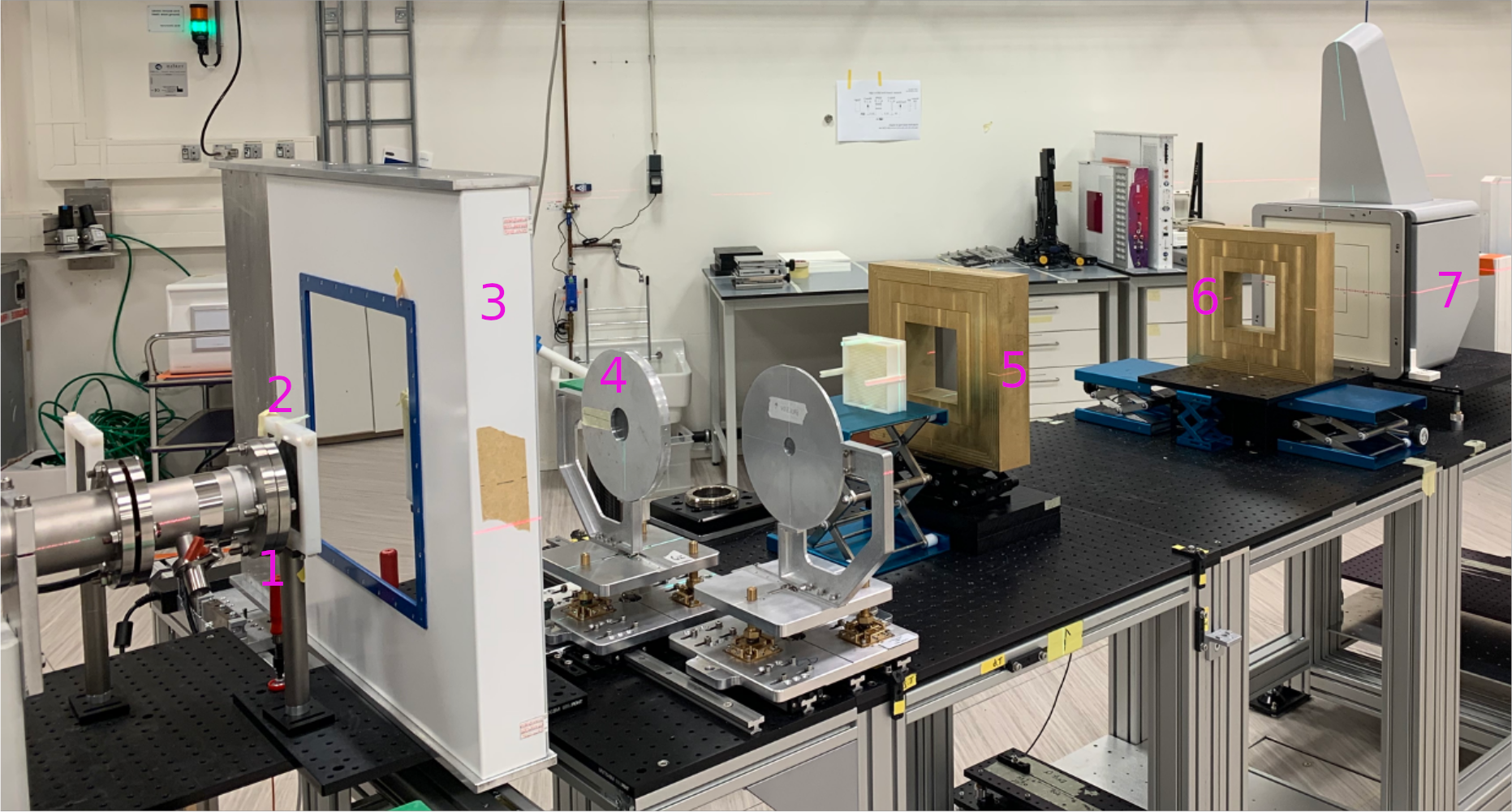}
    \includegraphics[width=0.70\textwidth]{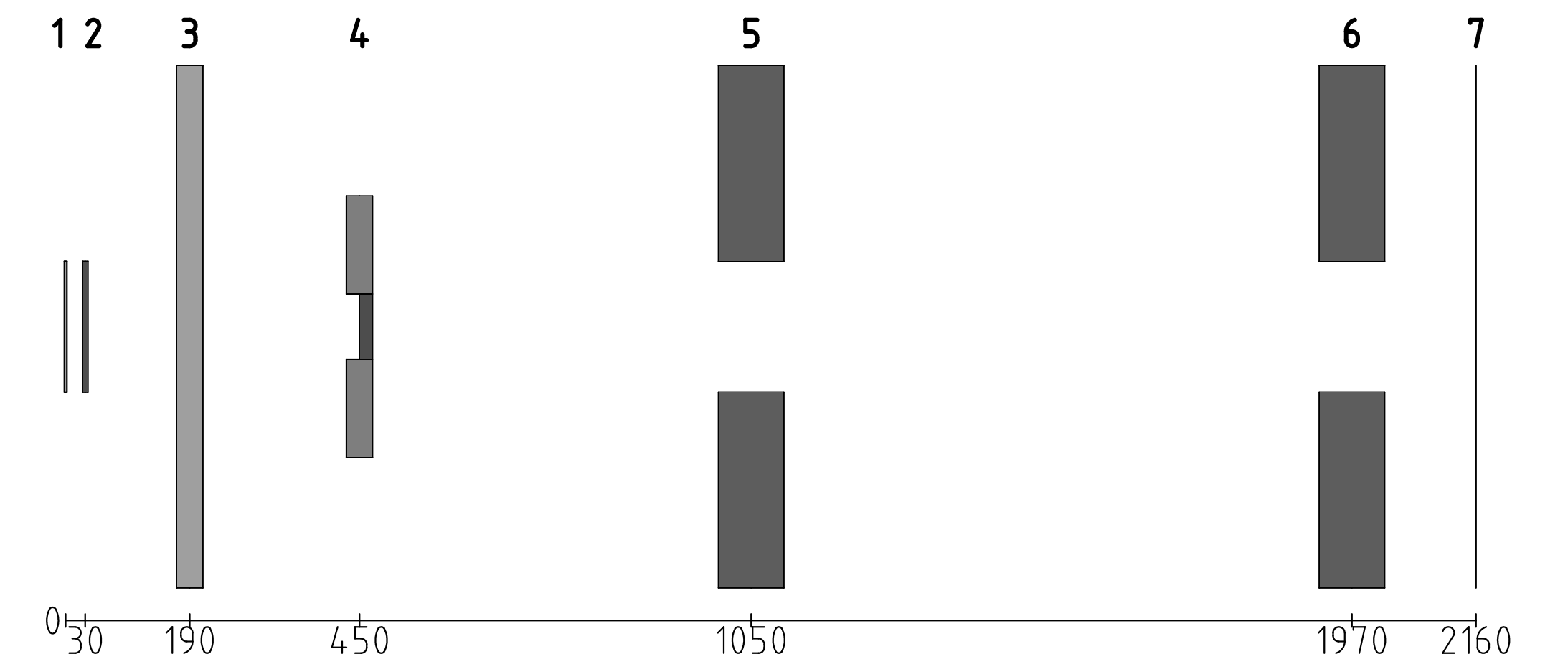}
    \caption{HollandPTC R\&D beamline configured in its passively scattered field configuration (top). Seven of the eight implemented geometric elements in G4HPTC-R\&D to mimic this experimental configuration, and their relative distances with respect to the Kapton vacuum pipe exit window in mm, can be seen in (bottom). Here the seven of the eight key geometric elements are: 1) the kapton vacuum pipe exit window, 2) scattering foil, 3) beam monitor, 4) dual ring, 5) first stage beam defining collimator, 6) second stage beam defining collimator, and 7) the front of the irradiation/measurement stage. Note that in (top) a Lynx\textsuperscript{\tiny\textregistered} detector can be seen at at the irradiation/measurement stage, and the QUBEnext detector is not shown.}
    \label{fig1}
\end{figure}

Geant4 version 10.06.p01 was utilised to develop the G4HPTC-R\&D simulation platform based on the experimental geometry of the HollandPTC R\&D beamline in its passively scattered field configuration seen in Figure \ref{fig1}. A total of eight key geometric elements outlined in Table \ref{tab1} were implemented in G4HPTC-R\&D that could be enabled or disabled depending on the beamline operational mode (i.e. pencil beam or passively scattered field configuration). The scattering foil and dual ring seen in Figure \ref{fig1} facilitate the expansion and shaping of the initial proton pencil beam to generate a uniform intensity proton field. The expanded and shaped beam is collimated at two stages along its path through the use of a first and second stage variable open cross-section brass collimator to produce square fields of up to 200 mm $\times$ 200 mm. At the irradiation/measurement stage where different radiobiological endstations are placed in Figure \ref{fig1}, a Lynx\textsuperscript{\tiny\textregistered} detector (IBA Dosimetry, Schwarzenbruck, Germany) can be seen which is used to assess the shape and quality of the proton field. Finally, an additional geometrical element that is not show in Figure \ref{fig1} was implemented: a water box that mimics the properties of a QUBEnext (DE.TEC.TOR, Turin, Italy) detector \footnote{The QUBEnext detector is a 128 multi-layer ionisation chamber with 2.34 mm thick detector planes/channels with an effective proton water equivalent thickness of 310 mm and possess a 127 mm $\times$ 127 mm sensitive cross-sectional area.}. Table \ref{tab1} outlines the dimensions and materials of the different elements, with additional information relating to their design and orientation along the beamline's path outlined in Rovituso et al. \cite{Rovituso2023}. 

\begin{figure}
    \centering
    \includegraphics[width=1.0\textwidth]{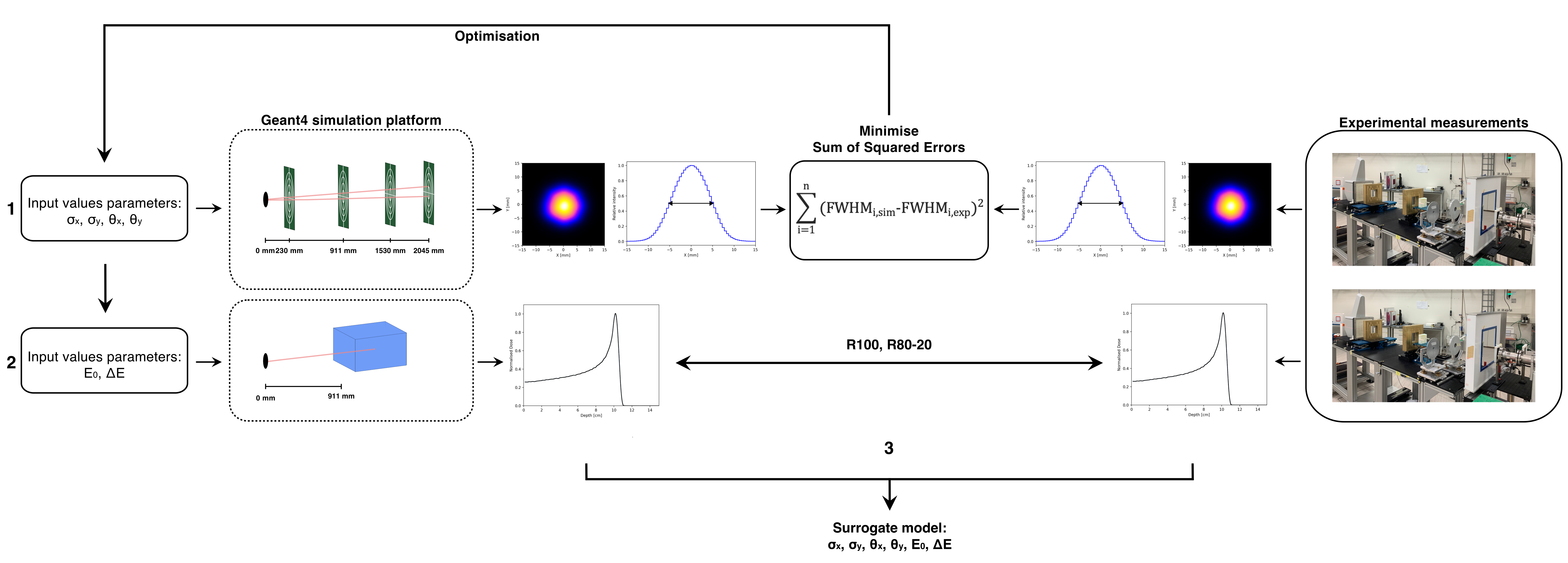}
    \caption{G4HPTC-R\&D simulation platform six parameter non-symmetrical Gaussian pencil beam surrogate model optimisation and individual parameter surrogate function mapping workflow with respect to experimental measurement at 70, 150, and 240 MeV. This parameter optimisation and individual surrogate function mapping workflow is composed of three stages: 1) lateral beam spot size ($\sigma_x$, $\sigma_y$) and angular deviation ($\theta_x$, $\theta_y$) optimisation with respect to the proton pencil beam envelope cross-sectionals evolution measurements in free air , 2) initial proton beam mean energy ($\textrm{E}_0$) and energy spread ($\Delta \textrm{E}$) optimisation with respect to proton pencil beam depth-dose profiles in water, and 3) mapping of each optimised parameter value to an individual parameter second-order polynomial surrogate function.}
    \label{fig2}
\end{figure}

The non-symmetrical Gaussian pencil proton beam of the HollandPTC R\&D beamline was implemented in G4HPTC-R\&D using a six parameter surrogate model emerging after the Kapton vacuum pipe exit window\footnote{Experimental characterisation of the HollandPTC R\&D beamline outlined in Rovituso et al. \cite{Rovituso2023} illustrated that its proton beam cross-section is ellipsoidal in nature and effectively horizontal along the length of its experimental range.}. These six parameters model three important characteristics of the proton beam at the Kapton vacuum pipe exit window: the 2D Gaussian lateral beam spot size ($\sigma_x$, $\sigma_y$), the 2D Gaussian beam spot angular deviation ($\theta_x$, $\theta_y$), and the 1D Gaussian energy spread of the cyclotron generated proton beam with initial mean energy $E_0$ and energy spread $\Delta E$. Through the optimisation workflow outlined in Figure \ref{fig2}, a set of individual surrogate functions were developed for each of these parameters to enable the modelling of the Gaussian pencil beam properties using the ProBeam isochronous cyclotron mean extraction proton energy with respect to experimental measurements at 70, 150, and 240 MeV. For each combination of these six parameters and detection media (Lynx\textsuperscript{\tiny\textregistered} detector or surrogate QUBEnext detector water box) investigated with G4HPTC-R\&D, a total of $10^6$ primary protons were run and the transport of all particles was simulated using a combined Geant4 ``Standard EM Option 4" and "QGSP\_BIC\_HP" physics list \cite{G42016,Arce2021} with atomic deexcitation enabled, a particle production range cut of 200 {\textmu}m, and a low-energy cut off of 250 eV.

In the first stage, the lateral beam spot size ($\sigma_x$, $\sigma_y$) and angular deviation ($\theta_x$, $\theta_y$) were optimised with respect to the proton pencil beam envelope cross-section measured with a Lynx\textsuperscript{\tiny\textregistered} detector in free air at at 230, 530, 911, 1230, 1530, 1830 and 2045 mm down-stream from the Kapton vacuum pipe exit window. Both sets of experimental and simulated Lynx\textsuperscript{\tiny\textregistered} detector data 2D beam profiles were fitted with a 2D Gaussian function to obtain the Full Width at Half Maximum (FWHM) in $x$ and $y$. The Full Width at Tenth Maximum (FWTM) was extracted from the central $x$- and $y$-axis planes to investigate the tails of the distributions as a second figure of merit. The agreement between experimental and simulated FWHM values as a function of the pencil beam energy was assessed through the us of the Sum of Squared Errors (SSE) metric:

\begin{equation}
\displaystyle \textrm{SSE} = \sum\limits_{i=1}^{n=7} (\textrm{FWHM}_{\textrm{sim,\textit{i}}} - \textrm{FWHM}_{\textrm{exp,\textit{i}}})^2 
\end{equation}

\noindent where $\textrm{FWHM}_{\textrm{sim,\textit{i}}}$ is the simulated and $\textrm{FWHM}_{\textrm{exp,\textit{i}}}$ is the experimental FWHM summed over $n=7$ distances. Starting with (predefined) initial values for $\sigma_x$, $\sigma_y$, $\theta_x$ and $\theta_y$ from Rovituso et al. \cite{Rovituso2023}, a two step optimisation with the SSE metric was undertaken. In the first step a $\pm10\%$ offset sweep around initial values was explored, and the combination that resulted in the smallest SSE was selected as an initial estimates for each parameter. A second parameter sweep of all four of the initial estimate for each parameter with a $\pm5\%$ offset was then undertaken to fine tune and ensure that each parameter value was not a local minimum in the optimisation.

In the second stage, the two remaining beam parameters ($E_0$ and $\Delta E$) that modelled the initial mean energy and energy spread of the cyclotron generated proton pencil beam were optimised with respect to the experimental proton pencil beam depth-dose profiles at 70, 150, and 240 MeV. Experimental measurements of the proton pencil beam depth-dose profiles at 70, 150, and 240 MeV were obtained with a QUBEnext detector placed at the beam isocentre 911 mm from the Kapton vacuum pipe exit window\footnote{The experimental 240 MeV proton pencil beam depth-dose profile was measured with a 100 mm thick and 300 mm $\times$ 300 mm cross-sectional area slab of water equivalent plastic placed in front of the QUBEnext detector.}. Using the optimised lateral beam spot size ($\sigma_x$, $\sigma_y$) and angular deviation ($\theta_x$, $\theta_y$) values at each energy, the mean energy ($E_0$) was varied in steps of 0.1 MeV around the ProBeam isochronous cyclotron mean extraction proton energy and the energy spread ($\Delta E$) in $\pm 5$ steps of 0.05 MeV with respect to initial estimates taken from Rovituso et al. \cite{Rovituso2023} to compare to the experimental proton pencil beam depth-dose profiles. Each proton pencil beam depth-dose profile data was fitted with a Bortfeld function \cite{Bortfeld1997}, and the position of the 80\% dose in the distal falloff (R80), the distance between the distal position of the 80\% and 20\% dose values (R80-R20 distal fall-off), and peak-to-entrance ratio were extracted as figures of merit \cite{Schuemann2014}. Comparison of the simulation and experimental results for these three figures of merit were used to optimise $E_0$, $\Delta E$, and to provided a general figure of merit.

The set of six parameter values obtained in the first and second stages of the optimisation workflow outlined in Figure \ref{fig2} form the basis of the developed individual parameter surrogate functions (third stage). They were solved through the mapping of each parameter value at 70, 150 and 240 MeV to second-order polynomial functions. With these surrogate functions G4HPTC-R\&D can model the non-symmetrical Gaussian pencil beam properties at the Kapton vacuum pipe exit window of the R\&D beamline with only the ProBeam isochronous cyclotron mean extraction proton energy as input.

\subsection{G4HPTC-R\&D Independent Experimental Benchmarking}

The refined non-symmetrical Gaussian pencil beam model and G4HPTC-R\&D simulation platform was benchmarked with respect to three independent experimental datasets. Two of these experimental datasets were of the HollandPTC R\&D beamline operating in its pencil beam configuration at 120 and 200 MeV, and the other was the HollandPTC R\&D beamline operating in its passively scattered field configuration at 150 MeV to generate a 100 mm $\times$ 100 mm field at the irradiation/measurement stage. Experimental measurements and G4HPTC-R\&D simulations were undertaken at 120 and 200 MeV in an identical manner to that outlined above to obtain beam envelope evolution in free air and depth-dose profiles in water datasets. These experimental and simulated FWHM, FWTM, R80, R80-R20 distal fall-off, and peak-to-entrance ratio results at each energy were compared to assess the validity of the refined non-symmetrical Gaussian pencil beam model and G4HPTC-R\&D simulation platform.

Experimental measurement of the 100 mm $\times$ 100 mm field generated at the irradiation/measurement stage for the HollandPTC R\&D beamline operating in its passively scattered field configuration was undertaken using the Lynx\textsuperscript{\tiny\textregistered} detector at 150 MeV. Large field simulations were also performed at 150 MeV with all beam elements implemented as shown in Figure \ref{fig1} and for an inner open cross-section of the final stage beam defining brass collimator set to 100 mm $\times$ 100 mm. A total of $4\times10^7$ primary protons were run, and the transport of all particles was simulated using the same physics configuration outlined above for the pencil beam configuration simulations. Three figures of merit were utilised to assess the validity of the refined G4HPTC-R\&D simulation platform operating in a passively scattered field configuration: 1 ) the field uniformity $U$, 2) the $\gamma$-index mean value, and 3) the $\gamma$-index global pass rate. The field uniformity ($U$) figure of merit across the field for both the experimental and simulation data was calculated by:

\begin{equation}
\displaystyle \textrm{U [\%]} = \left(1 - \frac{I_{max} - I_{min}}{I_{max} + I_{min}} \right) \cdot 100\% 
\end{equation}

\noindent where $I_{max}$ is the maximum intensity across the field, and $I_{min}$ is the minimum intensity across the field. It should be noted that a uniformity of above 97\% is required for radiobiological experiments \cite{Rovituso2023,Tommasino2019}. Whereas for the $\gamma$-index mean value and global pass rate figures of merit, the $\gamma$-index can be written:

\begin{equation}
\displaystyle \gamma(\mathbf{r_E}) = \min \{ \Gamma (\mathbf{r_E}, \mathbf{r_S}) \} \forall \{ \mathbf{r_S} \} 
\end{equation}

\noindent with: 

\begin{equation}
\displaystyle \Gamma (\mathbf{r_E}, \mathbf{r_S}) = \sqrt{\frac{\Delta r^{2} (\mathbf{r_E}, \mathbf{r_S})}{\delta r^{2}} + \frac{\Delta D^{2} (\mathbf{r_E}, \mathbf{r_S})}{\delta D^{2}}}
\end{equation}

\noindent where $\mathbf{r_E}$ is the spatial location in the experimental field, $\mathbf{r_S}$ is the spatial location in the simulated G4HPTC-R\&D field, $\Delta r (\mathbf{r_E}, \mathbf{r_S})$ is the distance between the two locations, $\Delta D (\mathbf{r_E}, \mathbf{r_S})$ difference in dose of between the two locations, $\delta r$ is the distance difference criterion, and $\delta D$ is the dose difference criterion \cite{Low1998,Hussein2017}. A total of three different $\delta r$ and $\delta D$ criterion combinations were assessed respectively: 1) 3 mm and 3\%, 2) 4 mm and 4\%, and 3) 5 mm and 5\%.


\section{Results and Discussion \label{sec:exp-results}}

\subsection{G4HPTC-R\&D Proton Pencil Beam Model Optimisation}

Table \ref{tab2} presents the optimal values for the non-symmetrical Gaussian pencil beam surrogate model parameters $\sigma_x$, $\sigma_y$, $\theta_x$ and $\theta_y$ obtained through the optimisation workflow outlined in Figure \ref{fig2} at 70, 150 and 240 MeV. In the case of $\sigma_x$, $\sigma_y$, $\theta_x$, $\theta_y$ and $\Delta \textrm{E}$, all of these parameters decrease with increasing initial mean proton energy as expected \cite{Paganetti2004,Fracchiolla2015,Resch2019}. Whereas for the $\textrm{E}_0$, the obtained values are within less than 2\% of the requested ProBeam isochronous cyclotron mean extraction proton energy. This small difference between experimental and G4HPTC-R\&D value for the $\textrm{E}_0$ parameter can be attributed to: 1) generation of the non-symmetrical Gaussian pencil beam spot after the Kapton vacuum pipe exit window, 2) uncertainties in the alignment and the resolution of the QUBEnext detector, and 3) uncertainties in relevant Geant4 proton cross-sectional data and physics models which are on the order of $\pm10\%$ \cite{Arce2021,Resch2019}. 

\begin{figure}
    \centering
    \subfloat[70 MeV]{\includegraphics[width=0.5\textwidth]{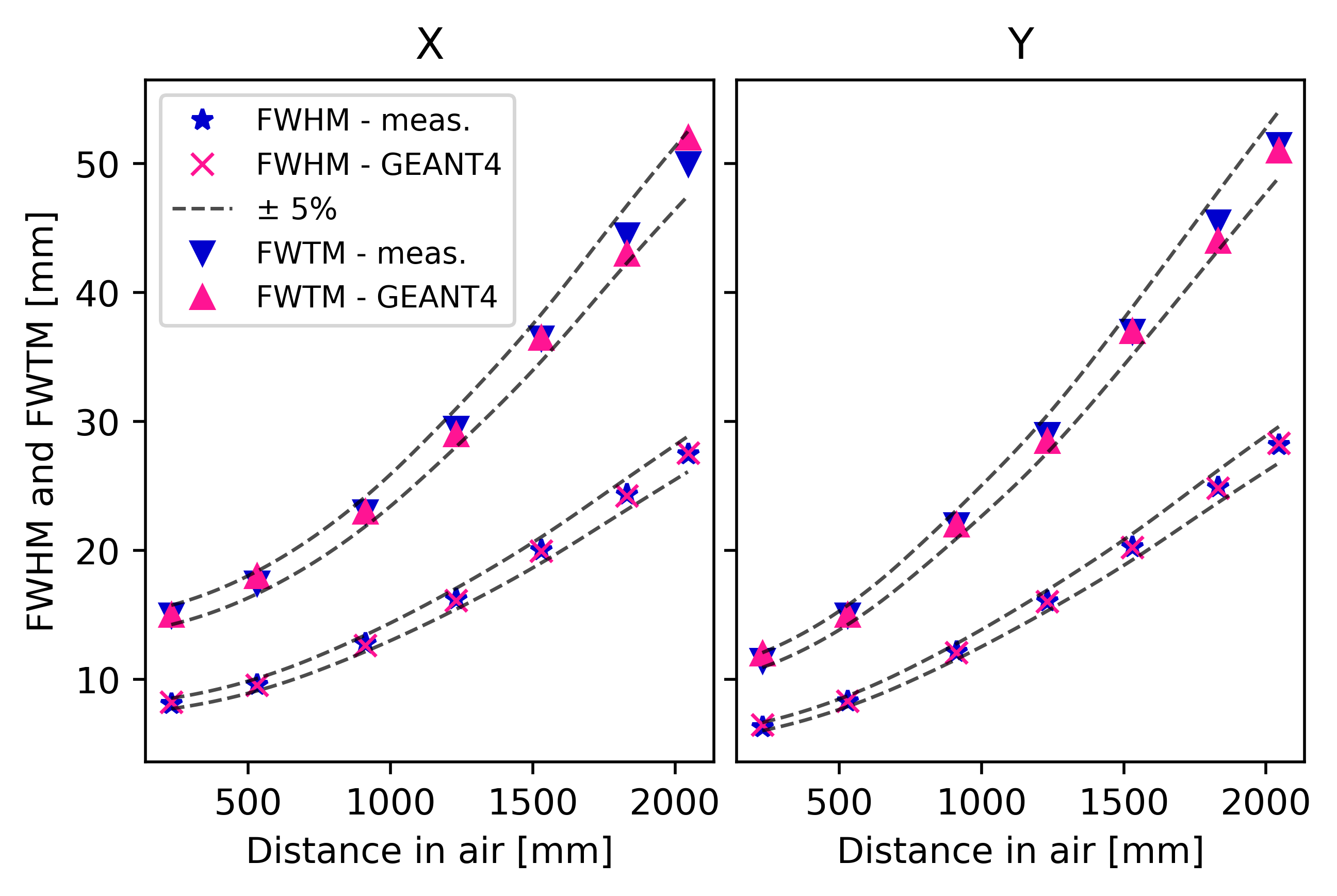}} \\
    \subfloat[150 MeV]{\includegraphics[width=0.5\textwidth]{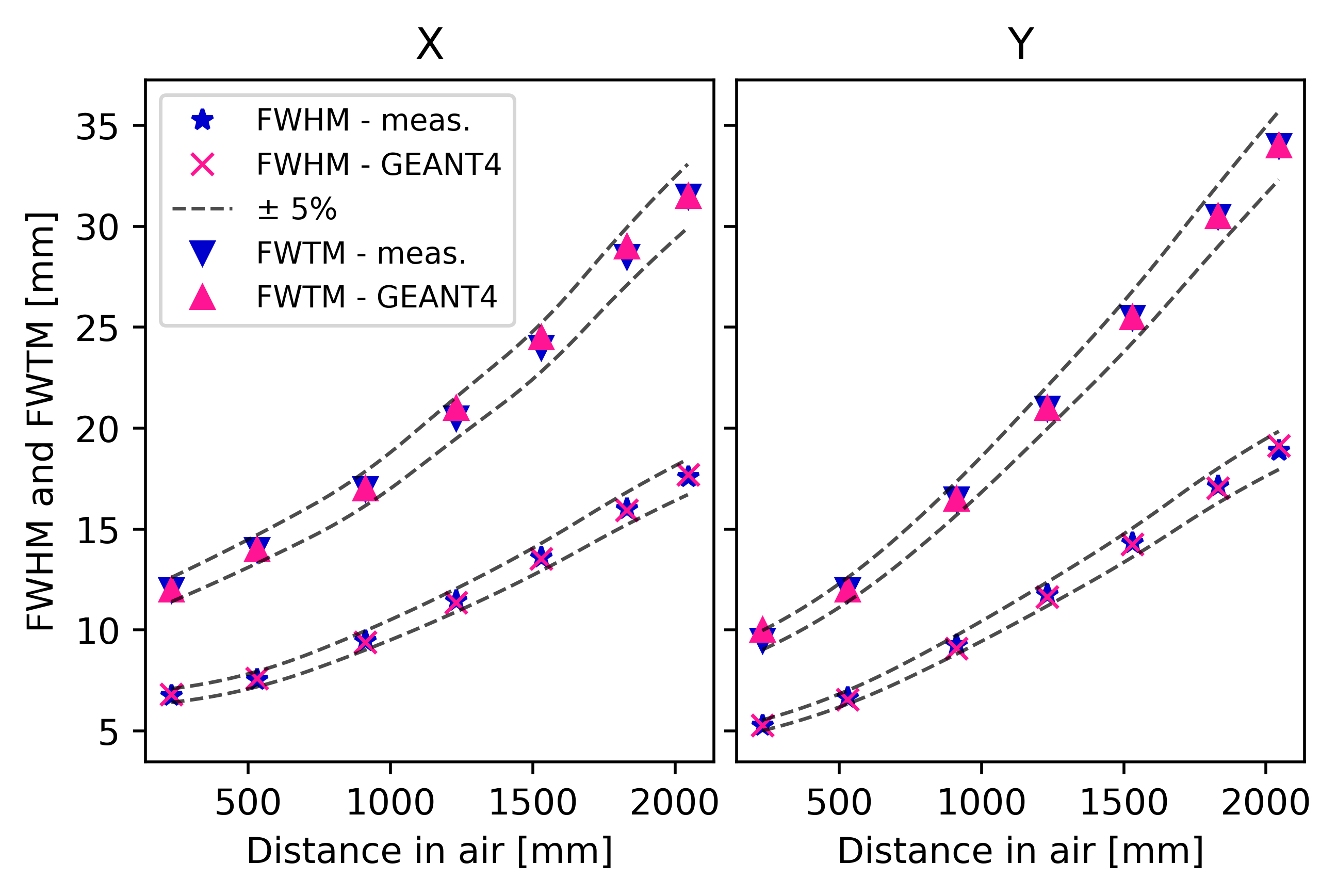}} \\
    \subfloat[240 MeV]{\includegraphics[width=0.5\textwidth]{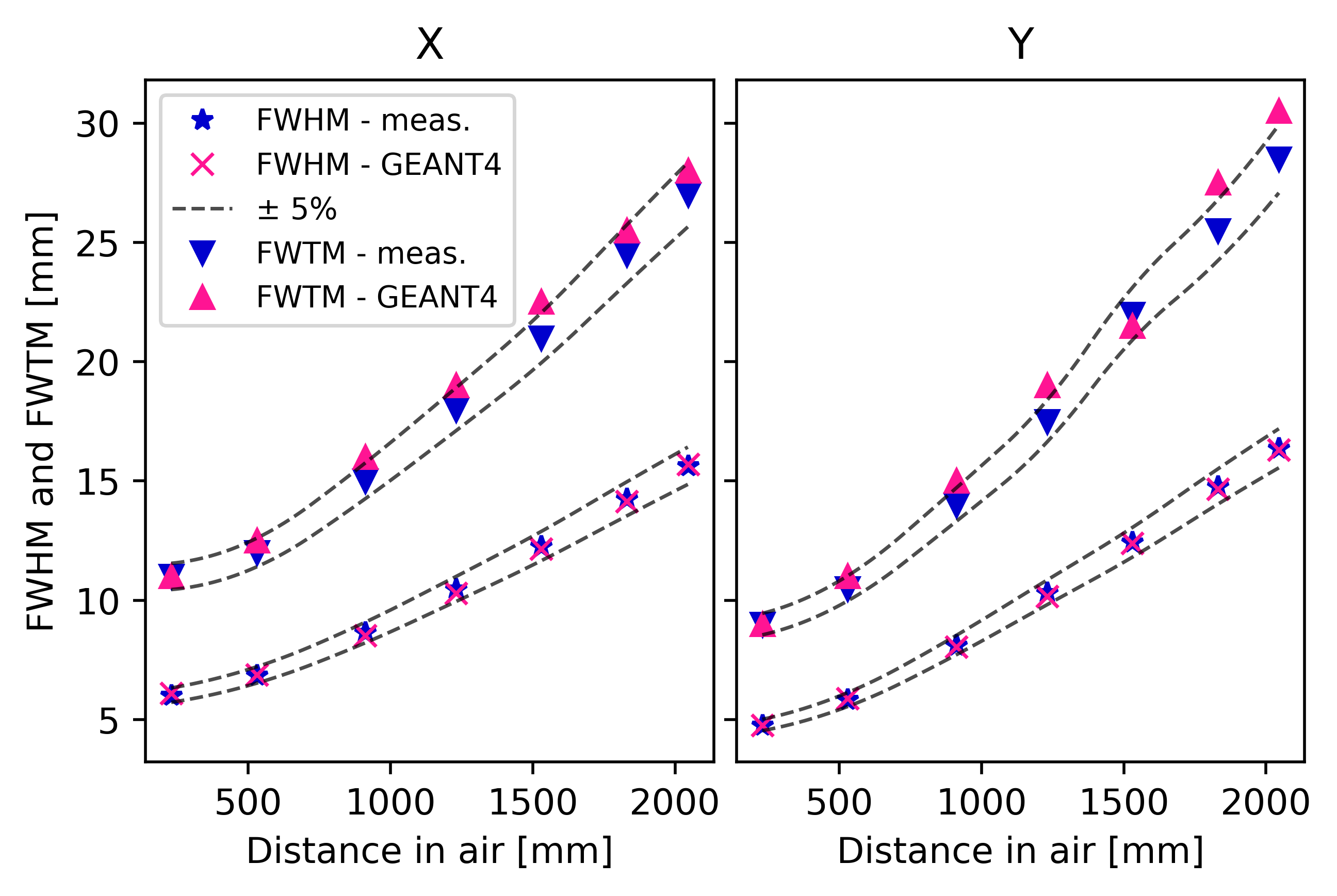}}
    \caption{Experimental and G4HPTC-R\&D 2D beam cross-section FWHM and FWTM values as a function of distance from the Kapton vacuum pipe exit window at 70, 150 and 240 MeV. Here error bars representing the accuracy of FHWM data fitting cannot be resolved due to their scale being on the same order as FWHM data symbols.}
    \label{fig4}
\end{figure}

Figure \ref{fig4} presents a comparison between the fitted FWHM and FWTM extracted values from the experimental and G4HPTC-R\&D proton pencil beam envelope cross-section measurements in free air at 70, 150, and 240 MeV using the optimal values for $\sigma_x$, $\sigma_y$, $\theta_x$, and $\theta_y$ outlined in Table \ref{tab2}. The dotted lines represent a $\pm$5\% deviation with respect to experimental results to aid in assessing the accuracy of the G4HPTC-R\&D results. For all energies the G4HPTC-R\&D FWHM values in both $x$ and $y$ were reproduced to within less than 5\% indicating a high level of correlation. With the FWTM values, the G4HPTC-R\&D 70 and 150 MeV data is also within 5\% of experimental results indicating a high level of correlation. However, at 240 MeV the G4HPTC-R\&D 2D proton pencil beam envelope cross-section FWTM values increase as a function of distance in air with the furthest two distances having a difference of up to 9\% relative to their respective experimental values. It should be noted that this level difference is still within the established uncertainties in relevant proton Geant4 proton cross-sectional data and physics models which are on the order of $\pm10\%$ \cite{Arce2021,Resch2019}, and the fact that this difference is still on the order of 5\% indicates a high level of correlation between the experimental and G4HPTC-R\&D data.

\begin{figure}
    \centering
    \includegraphics[width=0.8\textwidth]{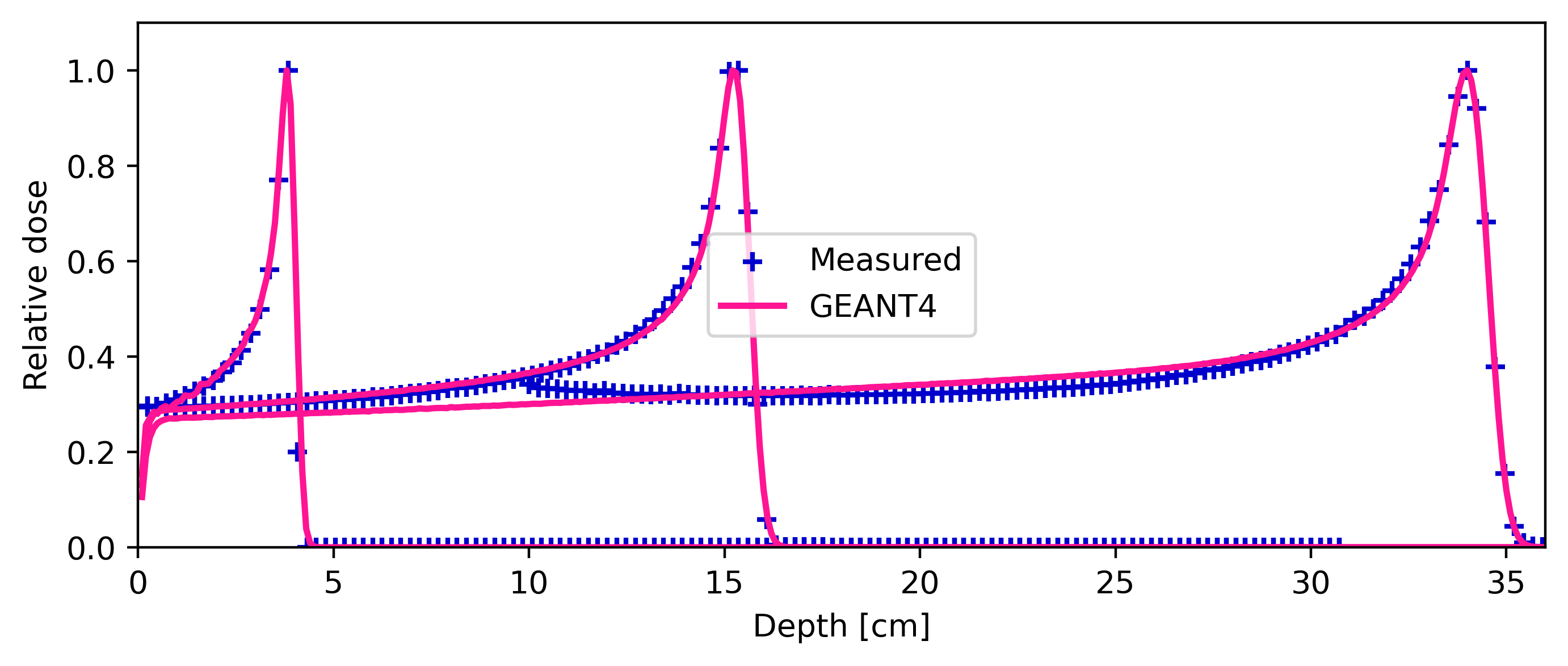}
    \caption{Experimental and G4HPTC-R\&D depth-dose profiles in water at 70, 150, 240 MeV.}
    \label{fig5}
\end{figure}

The depth-dose distributions of experimental and G4HPTC-R\&D results obtained with these optimised values of Table \ref{tab2} are shown in Figure \ref{fig5}, and their corresponding proton range (R80), distal fall-off (R80-R20) and peak-to-entrance ratio values are displayed in Table \ref{tab3}. The depth-dose profiles are normalised on the maximum dose, and show agreement in R80 and R80-R20 to within less than 0.5 mm (or 2\%) for all energies. As for the peak-to-entrance ratio, a difference on the order of 5\% can be observed for all three energies. Again given that the difference for all figures of merit are on the order or less than 5\%, a high level of correlation is present between the experimental and G4HPTC-R\&D depth-dose profiles at 70, 150, and 240 MeV.

\subsection{Refined G4HPTC-R\&D Proton Pencil Beam Model Independent Experimental Benchmarking}

\begin{figure}
    \centering
    \subfloat[120 MeV]{\includegraphics[width=0.5\textwidth]{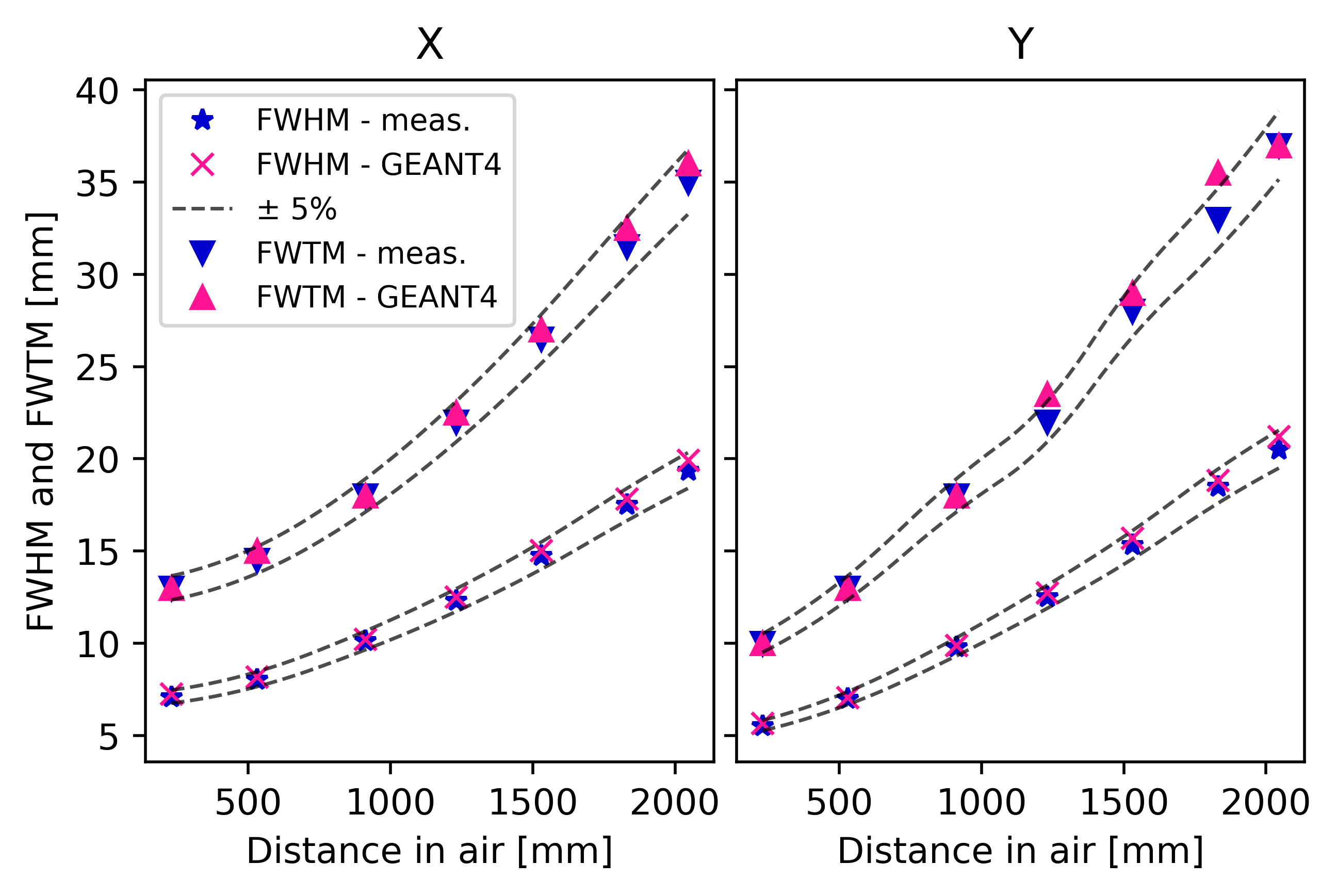}}
    \subfloat[200 MeV]{\includegraphics[width=0.5\textwidth]{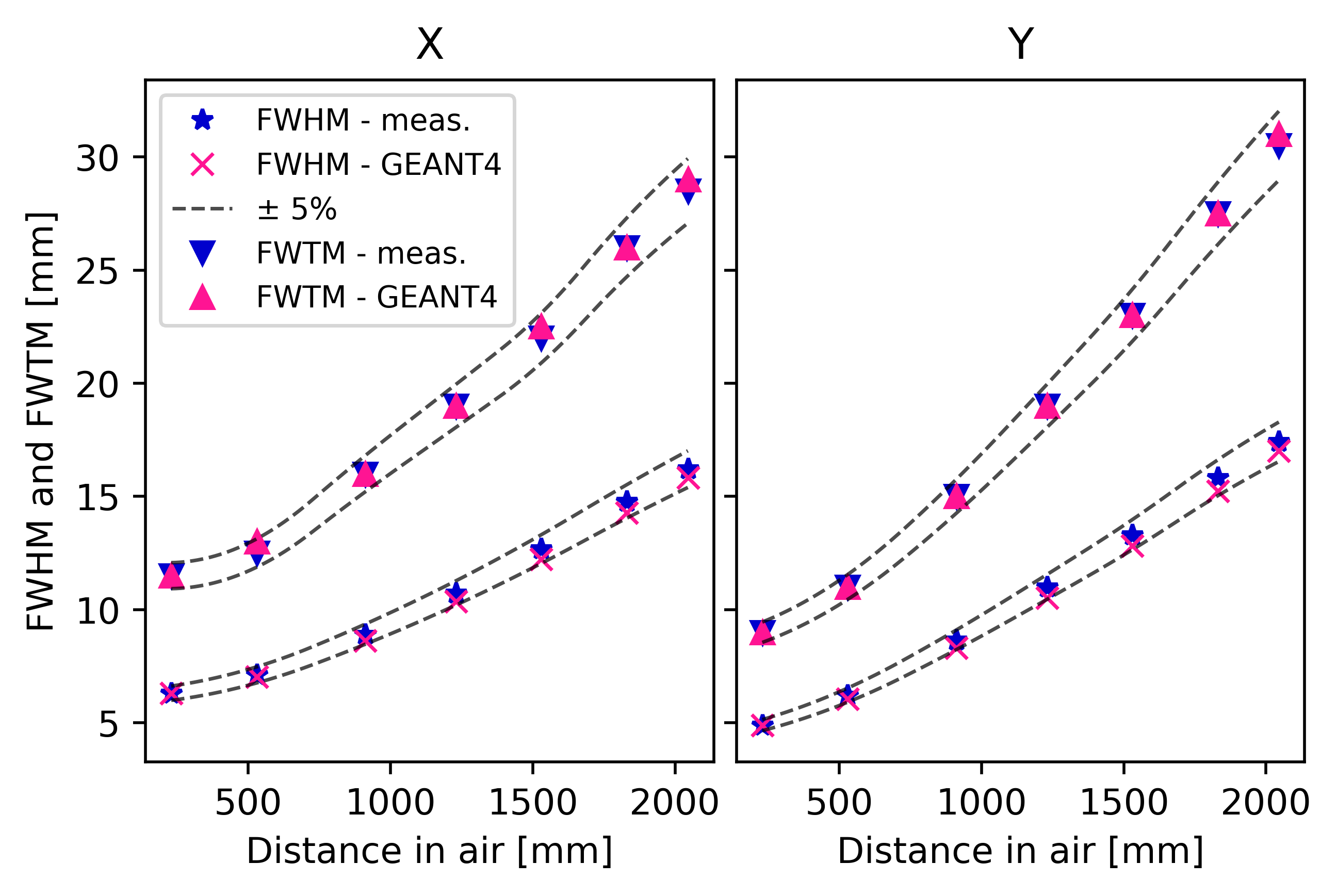}}
    \caption{Experimental and G4HPTC-R\&D 2D beam cross-section FWHM and FWTM values as a function of distance from the Kapton vacuum pipe exit window at 120, and 200 MeV. Here error bars representing the accuracy of FHWM data fitting cannot be resolved due to their scale being on the same order as FWHM data symbols.}
    \label{fig7}
\end{figure}

Table \ref{tab4} presents the second-order polynomial function constant values of the six refined non-symmetrical Gaussian pencil beam surrogate model parameters obtained via the optimisation workflow outlined in Section \ref{sec:exp-methods}. A comparison between the fitted FWHM and FWTM extracted values from the experimental and refined G4HPTC-R\&D proton pencil beam envelope cross-section measurements in free air for a mean ProBeam isochronous cyclotron extraction proton energy of 120 and 200 MeV using these values can be seen in Figure \ref{fig7}. As in Figure \ref{fig4}, the dotted lines represent a $\pm$5\% deviation with respect to experimental results to aid in assessing the accuracy of the G4HPTC-R\&D results. Inspection of Figure \ref{fig7} shows that the G4HPTC-R\&D FWHM values in both $x$ and $y$ were reproduced to within 5\% of the experimental data at 120 and 200 MeV. With the exception of a single outlier in the 120 MeV data at a distance of 1830 mm (7.6\% difference in the $y$), it can be seen that the G4HPTC-R\&D FWTM values in both $x$ and $y$ were reproduced again to within 5\% of the experimental data at the 120 and 200 MeV. 

\begin{figure}[]
    \centering
    \includegraphics[width=0.8\textwidth]{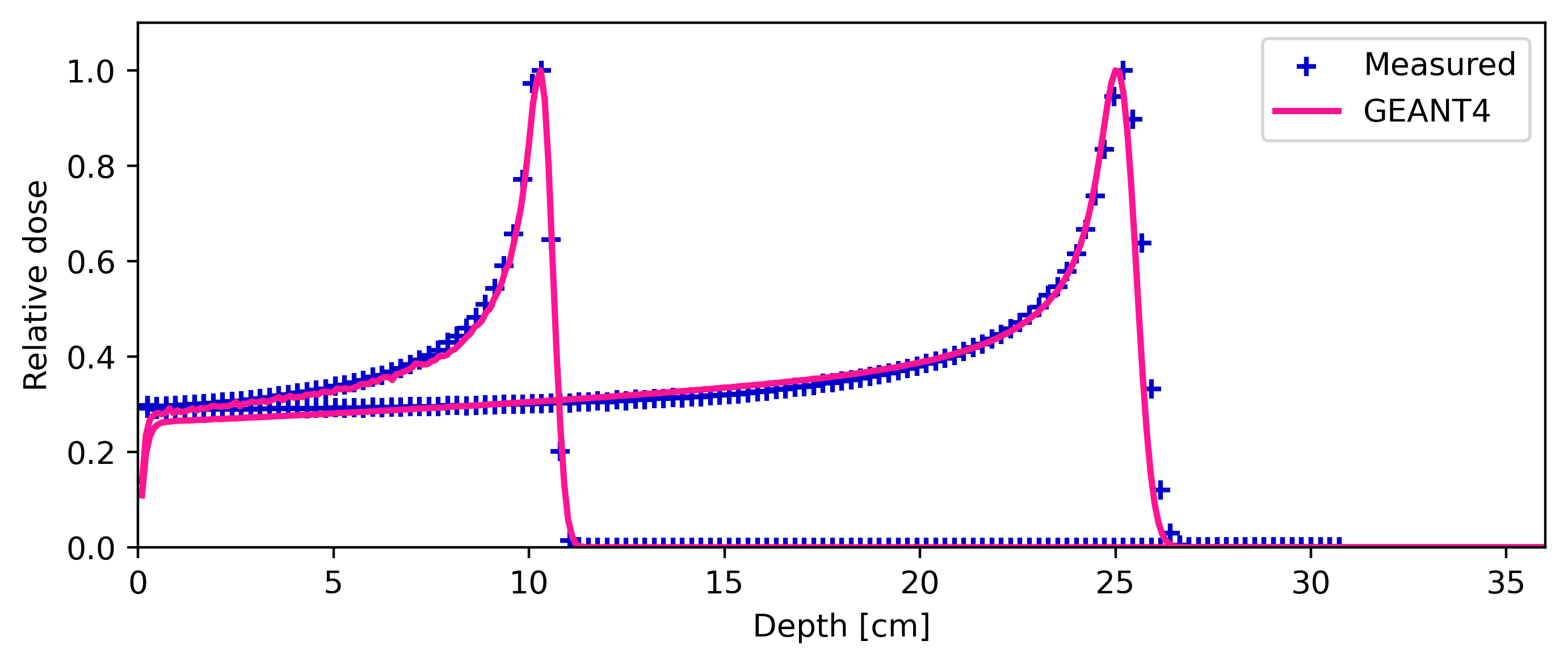}
    \caption{Experimental and G4HPTC-R\&D depth-dose profiles in water at 120 and 200 MeV.}
    \label{fig8}
\end{figure}

The depth-dose distributions of experimental and refined G4HPTC-R\&D results for a mean ProBeam isochronous cyclotron extraction proton energy of 120 and 200 MeV are shown in Figure \ref{fig8}, and their corresponding proton range (R80), distal fall-off (R80-R20) and peak-to-entrance ratio values are displayed in Table \ref{tab5}. For the 120 MeV depth-dose profile data the G4HPTC-R\&D R80 value agrees to within 0.5 mm (1\%) of the experimental data, and R80-R20 value is the same as the experimental data. Whereas for the 200 MeV depth-dose profiles, the G4HPTC-R\&D R80 value is within 1.5 mm (1\%) to the experimental data and R80-R20 is on the order of 0.5 mm (11\%) less then the experimental data. Furthermore, Table \ref{tab5} also shows that a difference on the order or less than 5\% is present in the experimental and G4HPTC-R\&D peak-to-entrance ratio values at both the 120 and 200 MeV. 

\begin{figure}
    \centering
    \includegraphics[width=0.8\textwidth]{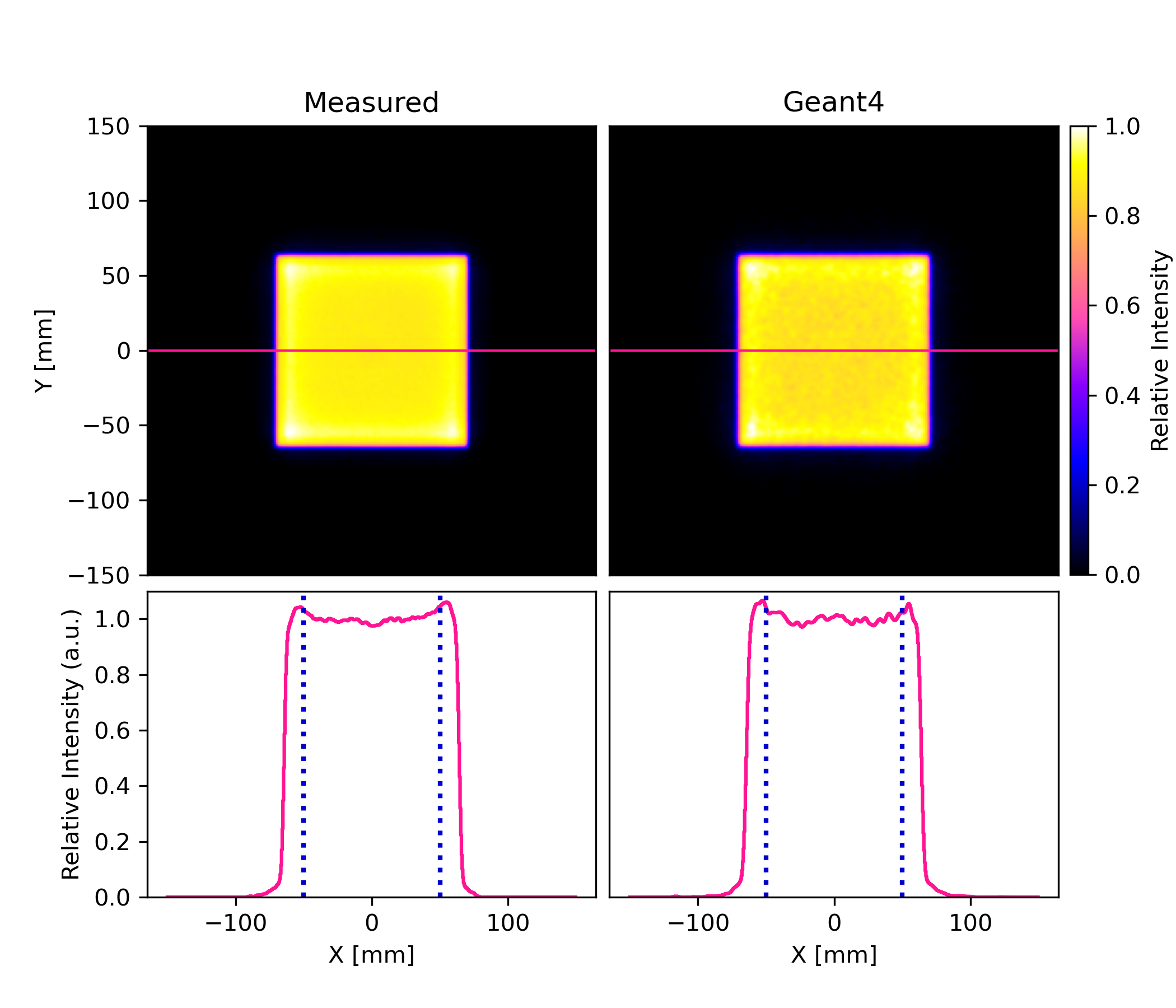}
    \caption{Experimental and G4HPTC-R\&D 100 mm $\times$ 100 mm fields (top) and their respective central $x$-axis lines profiles (bottom) generated at the irradiation/measurement stage of the HollandPTC R\&D beamline operating in its passively scattered field configuration. Here the pink horizontal lines (top) represent where the central $x$-axis lines profiles were taken, and the blue dotted lines outline the target field 100 mm $\times$ 100 mm of interest.}
    \label{fig9}
\end{figure}

Figure \ref{fig9} presents the experimental and refined G4HPTC-R\&D 100 mm $\times$ 100 mm fields and their respective central $x$-axis lines profiles generated at the irradiation/measurement stage of the HollandPTC R\&D beamline operating in its passively scattered field configuration. Both fields appear similar in shape and intensity, with their central $x$-axis lines profiles illustrating that the maximum difference in between the two is less than 5\%. Assessment of the field uniformity in the $x$-axis yields a value of 97.4\% and 96.7\%, and in the $y$-axis yields a value of 97.7\% and 96.9\% for the experimental and G4HPTC-R\&D data respectively. This less than 1\% field uniformity difference in along both axis indicates a good level of agreement between the experimental and G4HPTC-R\&D data, and that the field quality is sufficient for radiobiological experiments \cite{Rovituso2023,Tommasino2019}.

Table \ref{tab6} presents the G4HPTC-R\&D 100 mm $\times$ 100 mm field $\gamma$-index mean and $\gamma$-index global pass rate values with respect to the experimental 100 mm $\times$ 100 mm field for three different $\delta r$ and $\delta D$ criterion combinations. It can be seen that as the $\delta r$ and $\delta D$ combinations increase in value, the $\gamma$-index mean value decreases and the $\gamma$-index global pass rate increases. Under the most strict $\delta r$ = 3 mm and $\delta D$ = 3\% criterion combination, the $\gamma$-index global pass rate is 96.6\% and exceeds the clinically accepted pass rate threshold of 90\% for this criterion combination \cite{AAPM2009,Muller2019}. For $\delta r$ = 4 mm and $\delta D$ = 4\% the $\gamma$-index global pass rate increases to 99.9\%, and then to 100\% for $\delta r$ = 5 mm and $\delta D$ = 5\% criterion combination. 

These three independent experimental benchmarking trials of the refined G4HPTC-R\&D simulation platform illustrated that it is able to reproduce the physical characteristics of the HollandPTC R\&D beamline operating in both its pencil beam and passively scattered field configurations to within an acceptable level of agreement for clinical and radiobiological applications. The slight differences observed between the refined G4HPTC-R\&D and experimental data for these three independent experimental benchmarking trials can be attributed to two primary factors: 1) uncertainties in the alignment each beamline element and the resolution of the Lynx\textsuperscript{\tiny\textregistered}/QUBEnext detector, and 2) uncertainties in relevant Geant4 proton cross-sectional data and physics models which are on the order of $\pm10\%$ \cite{Arce2021,Resch2019}. The impact of beamline geometric element alignment uncertainties is particular relevant in the case of the 100 mm $\times$ 100 mm passively scattered field data where misalignment on the level of a millimetre or two can cause excessive "haloing" around the edges of the field. Even under optimal alignment conditions, this "haloing" effect is still present and can be observed in the both 2D field maps and central $x$-axis line profiles seen in Figure \ref{fig9} via the increased in relative intensity at the field edges and corners. Further work is already underway to explore the impact of these element alignment uncertainties on the 2D proton Linear Energy Transfer (LET) spectra distribution at the surface of cell/tissue culture for the different radiobiological endstations under development at the HollandPTC R\&D beamline \cite{Rovituso2023}.

\section{Conclusion \label{sec:diss-con}}

This work both developed and characterised the performance of a Geant4 based simulation platform of the Holland Proton Therapy Centre (Netherlands) R\&D beamline (G4HPTC-R\&D) to enable the planning, optimisation, and advanced dosimetry for radiobiological studies in both a pencil beam and passively scattered field configuration. It implemented a six parameter non-symmetrical Gaussian pencil beam surrogate model to simulate the R\&D beamline in both a pencil beam and passively scattered field configuration. Three different experimental proton datasets (70 MeV, 150 MeV, and 240 MeV) of the pencil beam envelope evolution in free air and depth-dose profiles in water were used to develop a set of individual parameter surrogate functions to enable the modelling of the non-symmetrical Gaussian pencil beam properties with only the ProBeam isochronous cyclotron mean extraction proton energy as input. This refined beam model was then benchmarked with respect to three independent experimental datasets of the R\&D beamline operating in both a pencil beam configuration at 120 and 200 MeV, and passively scattered field configuration at 150 MeV. It was shown that the G4HPTC-R\&D simulation platform can reproduce the pencil beam envelope evolution in free air and depth-dose profiles to within an accuracy on the order of $\pm$5\% for all tested energies, and that it was able to reproduce the 150 MeV passively scattered field to the specifications need for clinical and radiobiological applications.

\section*{Acknowledgements}

This work was partially funded by Varian Medical Systems (a Siemens Healthineers Company), and the simulations undertaken on the Dutch national e-infrastructure with the support of SURF Cooperative (Grant No: EINF-486 (2020)). J.~M.~C.~Brown was supported by a Veni fellowship from the Dutch Organization for Scientific Research (NWO Domain AES Veni 16808 (2018)) during the time of this study.


\newpage


\newpage

\begin{table}[]
\centering
\begin{tabular}{ccc}
\hline
\textbf{Name} & \textbf{Dimensions} & \textbf{Material} \\
\hline
 Kapton Vacuum  & Disk (diameter,z):  & G4\_KAPTON \\
  Pipe Exit Window & 100, 0.125 mm & \\
   \hline
 Scattering Foil  & Box (x,y,z):  & G4\_Pb \\
  & 100, 100, 1.7 mm & \\
   \hline
  Beam Monitor  & Surrogate Box (x,y,z):  & G4\_WATER \\
  & 400, 400, 0.6 mm & \\   
  \hline
 Dual  & Inner Disk (diameter,z):  & G4\_Pb \\
 Ring & 45, 5.5 mm & \\   
  & Outer Ring (inn., out., z):  & G4\_Al \\
  & 45, 200, 16 mm & \\   
   \hline
 First Stage & Box Outer (x,y,z):  & Brass \\
 Collimator & 400, 400, 50 mm & Cu:Zn:Pb \\   
 & Inner Opening (x,y,z):  & 58\%:39\%:3\% \\
 & 100, 100, 50  mm & $\rho=$8.7 g/cm$^3$ \\   
   \hline
 Second Stage & Box Outer (x,y,z):  & Brass \\
 Collimator & 400, 400, 50 mm & Cu:Zn:Pb \\   
 & Inner Opening (x,y,z):  & 58\%:39\%:3\% \\
 & 100, 100, 50  mm & $\rho=$8.7 g/cm$^3$ \\   
   \hline
  Lynx\textsuperscript{\tiny\textregistered} Detector & Front Box (x,y,z):  & G4\_PLEXIGLASS \\
   & 300, 300, 1 mm & \\   
   & Back Box (x,y,z):  & G4\_GADOLINIUM  \\
   & 300, 300, 0.5 mm &  \_OXYSULFIDE \\
   \hline
  Water Box & Box (x,y,z):  & G4\_WATER \\
  (QUBEnext Detector) & 127, 127, 400 mm & \\
   \hline   
\end{tabular}%

\caption{Name, dimensions, and materials of the geometric elements that were implemented in G4HPTC-R\&D to mimic the experimental configuration seen in Figure \ref{fig1}.}
\label{tab1}
\end{table}

\begin{table}[]
\centering
\begin{tabular}{ccccccc}
\hline
\multicolumn{1}{l}{E (MeV)} & $\sigma_x$ (mm) & $\sigma_y$ (mm) & $\theta_x$ (rad) & $\theta_y$ (rad) & $\textrm{E}_0$ (MeV) & $\Delta \textrm{E}$ (MeV)\\ \hline
70  & 3.383 & 2.559 & 0.00371 & 0.00409 & 69.8    & 1.43 \\ 
150 & 2.819 & 2.100 & 0.00280 & 0.00330 & 148.8   & 1.35 \\ 
240 & 2.509 & 1.890 & 0.00273 & 0.00300 & 235.5   & 0.8 \\ 
\hline
\end{tabular}
\caption{Optimised non-symmetrical Gaussian pencil beam surrogate model parameters at 70, 150 and 240 MeV: lateral spread $\sigma$ in $x$ and $y$ (mm) and angular spread $\theta$ in $x$ and $y$ (radians), initial mean energy $\textrm{E}_0$ (MeV) and energy spread $\Delta \textrm{E}$ (MeV).}
\label{tab2}
\end{table}

\begin{table}
\centering
\begin{tabular}{ccccccc}
\hline
E (MeV) & \multicolumn{2}{l}{R80 (mm)} & \multicolumn{2}{l}{R80-R20 (mm)} & \multicolumn{2}{l}{Peak-Entrance ratio} \\
            & Exp.                 & Geant4                & Exp.                    & Geant4                    & Exp.              & Geant4             \\ \hline
70          & 39.1                  & 39.6                  & 1.6                     & 2.1                      & 3.4              & 3.6               \\
150         & 155.2                 & 155.2                 & 3.8                     & 3.8                      & 3.4               & 3.4               \\
240         & 343.6                  & 343.8                  & 5.4                     & 5.0                      & 3.2               & 3.4               \\ \hline
\end{tabular}
\caption{Proton range R80 (mm), distal fall-off R80-R20 (mm) and peak-to-entrance ratios of the experimental and G4HPTC-R\&D of the 70, 150, and 240 MeV depth-dose distributions seen in Figure \ref{fig5}. }
\label{tab3}
\end{table}

\begin{table}
\centering
\begin{tabular}{cccc}
\hline
Parameter & $A$ & $B$ & $C$ \\
\hline
$\sigma_x$ (mm) & $2.121 \times 10^{-5}$& $-1.172 \times 10^{-2}$ & 4.099 \\
$\sigma_y$ (mm) & $2.003 \times 10^{-5}$ & $-1.014 \times 10^{-2}$ & 3.171 \\
$\theta_x$ (rad) & $6.234 \times 10^{-5}$  & $-2.509 \times 10^{-2}$ & 5.161 \\
$\theta_y$ (rad) & $3.848 \times 10^{-5}$  & $-1.834 \times 10^{-2}$ & 5.185 \\
$\textrm{E}_0$ (MeV) & $-1.561 \times 10^{-4}$ & $1.023$ & $-1.151$ \\
$\Delta \textrm{E}$ (MeV) & $2.043 \times 10^{-5}$ & $-3.574 \times 10^{-3}$ & 1.840 \\
\hline
\end{tabular}
\caption{The second-order polynomial function constant values of the six refined non-symmetrical Gaussian pencil beam surrogate model parameters obtained via the optimisation workflow outlined in Section \ref{sec:exp-methods}. Here the second-order polynomial function is written $\textrm{Parameter (E)} = A \textrm{E}^{2} + B \textrm{E} + C$ where $\textrm{E}$ is the  mean extraction proton energy of the ProBeam isochronous cyclotron.}
\label{tab4}
\end{table}

\begin{table}
\centering

\begin{tabular}{ccccccc}
\hline
E (MeV) & \multicolumn{2}{l}{R80 (mm)} & \multicolumn{2}{l}{R80-R20 (mm)} & \multicolumn{2}{l}{Peak-Entrance ratio} \\
            & Meas.                 & Geant4                & Meas.                    & Geant4                    & Meas.              & Geant4             \\ \hline
120         & 104.7                 & 105.0                 & 3.3                     & 3.3                      & 3.4               & 3.6               \\
200         & 255.4                 & 253.9                 & 5.1                     & 4.5                      & 3.5               & 3.6               \\ \hline
\end{tabular}%
\caption{Proton range R80 (mm), distal fall-off R80-R20 (mm) and peak-to-entrance ratios of the experimental and G4HPTC-R\&D of the 120 and 200 MeV depth-dose distributions seen in Figure \ref{fig8}.}
\label{tab5}
\end{table}

\begin{table}[]
\centering
\begin{tabular}{ccc}
\hline
  $\delta r$ (mm), $\delta D$ (\%) & $\gamma$-Index Mean & $\gamma$-Index Global Pass Rate \\
\hline   

  3, 3 & 0.49 & 96.6 \\
  4, 4 & 0.36 & 99.9 \\
  5, 5 & 0.29 & 100.0 \\
\hline
\end{tabular}%

\caption{G4HPTC-R\&D 100 mm $\times$ 100 mm field $\gamma$-index mean and $\gamma$-index global pass rate values with respect to the experimental 100 mm $\times$ 100 mm field for three different $\delta r$ and $\delta D$ criterion combinations.}
\label{tab6}
\end{table}

\end{document}